\documentclass[aip,reprint,amsmath,amssymb,superscriptaddress]{revtex4-1}

\usepackage{verbatim} 

\usepackage{graphicx}
\usepackage{dcolumn}

\usepackage{amstext,amsmath,amssymb,amsfonts}
\usepackage{bm}

\usepackage{txfonts}
\usepackage{mathtools}
\usepackage{caption}
\usepackage{subcaption}
\usepackage{ulem}
\DeclareMathAlphabet{\mathpzc}{OT1}{pzc}{m}{it}
\DeclareFontFamily{OT1}{pzc}{}
\DeclareFontShape{OT1}{pzc}{m}{it}{<-> s * [1.100] pzcmi7t}{}
\DeclareMathAlphabet{\mathpzc}{OT1}{pzc}{m}{it}

\usepackage{hyperref}

\usepackage{color}
\definecolor{lightblue}{rgb}{0.2,0.2,0.7}
\definecolor{darkblue}{rgb}{0,0.25,0.5}
\definecolor{redbrown}{rgb}{0.875,0.25,0.125}
\definecolor{darkgreen}{rgb}{0,0.5,0}

\newcommand{\bra}[1]{\ensuremath{\langle #1 \vert}}
\newcommand{\ket}[1]{\ensuremath{\vert #1  \rangle}}
\newcommand{\braket}[2]{\ensuremath{\langle  #1 \vert #2  \rangle}}
\renewcommand{\b}[1]{\ensuremath{\mathbf{#1}}}

\newcommand{\LDA}{\ensuremath{\text{LDA}}}

\renewcommand{\d}{\ensuremath{\text{d}}}

\newcommand{\stat}[1]{\underset{#1}{\text{stat}}}

\renewcommand{\i}{\ensuremath{\text{i}}}
\newcommand{\B}{\ensuremath{{\cal B}}}
\newcommand{\FCI}{\ensuremath{\text{FCI}}}

\newcommand{\calh}{\ensuremath{\mathpzc{h}}}

\begin{document}

\title{Basis-set correction based on density-functional theory: Linear-response formalism for excited-state energies}

\author{Diata Traore}
\email{dtraore@lct.jussieu.fr}
\affiliation{Laboratoire de Chimie Théorique, Sorbonne Université and CNRS, F-75005 Paris, France}

\author{Emmanuel Giner}
\email{emmanuel.giner@lct.jussieu.fr}
\affiliation{Laboratoire de Chimie Théorique, Sorbonne Université and CNRS, F-75005 Paris, France}

\author{Julien Toulouse}
\email{toulouse@lct.jussieu.fr}
\affiliation{Laboratoire de Chimie Théorique, Sorbonne Université and CNRS, F-75005 Paris, France}
\affiliation{Institut Universitaire de France, F-75005 Paris, France}

\date{May 25, 2023}

\begin{abstract}
The basis-set correction method based on density-functional theory consists in correcting the energy calculated by a wave-function method with a given basis set by a density functional. This basis-set correction density functional incorporates the short-range electron correlation effects missing in the basis set. This results in accelerated basis convergences of ground-state energies to the complete-basis-set limit. In this work, we extend the basis-set correction method to a linear-response formalism for calculating excited-state energies. We give the general linear-response equations, as well as the more specific equations for configuration-interaction wave functions. As a proof of concept, we apply this approach to the calculations of excited-state energies in a one-dimensional two-electron model system with harmonic potential and a Dirac-delta electron-electron interaction. The results obtained with full-configuration-interaction wave functions expanded in a basis of Hermite functions and a local-density-approximation basis-set correction functional show that the present approach does not help in accelerating the basis convergence of excitation energies. However, we show that it significantly accelerates basis convergences of excited-state total energies.
\end{abstract}

\maketitle

\section{Introduction}

One of the main limitations of standard electronic-structure wave-function computational methods is their slow convergence of ground- and excited-state energies and other properties with respect to the one-electron basis set (see, e.g., Refs.~\onlinecite{HelKloKocNog-JCP-97,HalHelJorKloKocOlsWil-CPL-98,HelJorOls-BOOK-02,SheGruBooKreAla-PRB-12}). This slow convergence can be traced back to the non-smoothness of the exact eigenfunctions of the Schr\"odinger Hamiltonian with repulsive Coulomb electron-electron interaction~\cite{KutMor-JCP-92}, namely the electron-electron cusp condition~\cite{Kat-CPAM-57,PacBye-JCP-66}.

There are two main approaches for dealing with this slow basis convergence problem. The first approach consists in extrapolating the results to the complete-basis-set (CBS) limit by using increasingly large basis sets~\cite{HelKloKocNog-JCP-97,HalHelJorKloKocOlsWil-CPL-98}. This approach is very common for estimating the CBS limit of the ground-state energy but has also been used for estimating the CBS limit of excited-state energies and properties (see, e.g., Refs.~\onlinecite{KalGau-JCP-04,WatCha-JCTC-12,BauSanGarSan-JCP-14,ChrBloLooJac-JCTC-21}). The second approach consists in using explicitly correlated R12 or F12 methods which incorporate in the wave function a correlation factor reproducing the electron-electron cusp (see, e.g., Refs.~\onlinecite{TenNog-WIRES-12,HatKloKohTew-CR-12,KonBisVal-CR-12,ShiWer-MP-13}). The vast majority of R12/F12 methods have been applied to ground-state energy calculations but linear-response extensions have also been proposed for excitation energies and dynamic response properties~\cite{FliHatKlo-JCP-06,NeiHatKlo-JCP-06,NeiHat-JCP-07,Koh-JCP-09,HanKoh-JCP-09}.

Recently, some of the present authors introduced an alternative basis-set correction method based on density-functional theory (DFT)~\cite{GinPraFerAssSavTou-JCP-18}. It consists in correcting the energy calculated by a wave-function method (such as configuration interaction or coupled-cluster) with a given basis set by an adapted basis-set correction density functional incorporating the short-range electron correlation effects missing in the basis set, resulting in an accelerated convergence to the CBS limit. This basis-set correction method was further developed and validated on atomization energies~\cite{LooPraSceTouGin-JPCL-19,YaoGinLiTouUmr-JCP-20,YaoGinAndTouUmr-JCP-21} and dissociation energy curves~\cite{GinSceLooTou-JCP-20}. The method was also extended to calculations of ionization potentials within the GW approach~\cite{LooPraSceGinTou-JCTC-20} and to calculations of dipole moments~\cite{GinTraPraTou-JCP-21,TraTouGin-JCP-22}. It was also proposed to extend the method to calculations of excitation energies using a straightforward state-specific approach in which the same basis-set correction functional is evaluated from the density of each state~\cite{GinSceTouLoo-JCP-19}. Even though the last approach was shown to be able to accelerate the basis convergence of electronic excitation energies of molecular systems, it is based on the a-priori questionable assumption that one can use the same basis-set correction functional for all states. 

In the present work, we extend the basis-set correction method to a linear-response formalism, providing a more rigorous framework for calculating excitation energies. Moreover, it allows for calculations of response properties such as dynamic polarizabilities. As a first proof of concept, we apply this approach to calculations of excitation energies in a one-dimensional (1D) model system consisting of two electrons in a harmonic potential with a Dirac-delta two-electron interaction~\cite{MagBur-PRA-04,MaiZhaCavBur-JCP-04}. We previously used a similar 1D model system in Ref.~\onlinecite{TraGinTou-JCP-22} to study with some mathematical rigor the basis-set correction method. The relevance of this 1D model for quantum chemistry lies in the fact that the Dirac-delta two-electron interaction induces a slow basis convergence quite similar to the one observed with the standard two-electron Coulomb interaction in three-dimensional (3D) systems.

The paper is organized as follows. In Sec.~\ref{sec:linearresponse}, we formulate the general linear-response theory for the DFT-based basis-set correction scheme, and we give explicit expressions for configuration-interaction wave functions. In Sec.~\ref{sec:1Dmodel}, we apply the linear-response DFT-based basis-set correction theory to the 1D model system and we discuss the results. Finally, Sec.~\ref{sec:conclusions} contains our conclusions. Hartree atomic units are used throughout this work.

\section{Linear-response DFT-based basis-set correction}
\label{sec:linearresponse}

In this section, we derive the general linear-response equations for the DFT-based basis-set correction approach. We consider a finite one-electron basis set $\B \subset H^1(\mathbb{R}^3\times\{\uparrow,\downarrow\},\mathbb{C})$ where $H^1$ is the first-order Sobolev space. The corresponding one-electron Hilbert space spanned by this basis set is denoted by ${\calh}^\B = \text{span}(\B)$ and the corresponding $N$-electron Hilbert space is given by the $N$-fold antisymmetric tensor product of ${\calh}^\B$, i.e. ${\cal H}^\B = \bigwedge^N {\calh}^\B$. 

\subsection{General ground-state optimization}

We consider a general parametrized wave function $\ket{\Psi(\b{p})} \in {\cal H}^\B$ with $M$ complex-valued parameters $\b{p}=(p_1,p_2,...,p_M) \in \mathbb{C}^M$. For example, these parameters could be configuration-interaction coefficients or orbital-rotation parameters. For convenience, we work with the intermediately normalized wave function (see, e.g., Ref.~\onlinecite{MusCocAssOttUmrTou-AQC-18})
\begin{equation}
\ket{\bar{\Psi}(\b{p})} = \frac{\ket{\Psi(\b{p})}}{\braket{\Psi_0}{\Psi(\b{p})}},
\end{equation}
where $\ket{\Psi_0} = \ket{\Psi(\b{p}^0)}$ is the current wave function obtained for the current parameters $\b{p}=\b{p}^0$. The current wave function is taken as normalized to unity, i.e. $\braket{\Psi_0}{\Psi_0}=1$. The advantage of this intermediate normalization is that the first- and second-order derivatives of $\ket{\bar{\Psi}(\b{p})}$ with respect to $\b{p}$ at $\b{p}=\b{p}^0$,
\begin{equation}
\ket{\bar{\Psi}_I} = \left. \frac{\partial \ket{\bar{\Psi}(\b{p})}}{\partial p_I} \right|_{\b{p}=\b{p}^0}\;  \text{and} \;\;
\ket{\bar{\Psi}_{I,J}} = \left. \frac{\partial^2 \ket{\bar{\Psi}(\b{p})}}{\partial p_I \partial p_J} \right|_{\b{p}=\b{p}^0},
\label{PsiIandPsiIJ}
\end{equation}
are orthogonal to $\ket{\Psi_0}$, i.e. $\braket{\bar{\Psi}_I}{\Psi_0}=0$ and $\braket{\bar{\Psi}_{I,J}}{\Psi_0}=0$. This simplifies the derivation of the equations.

In the DFT-based basis-set correction approach~\cite{GinPraFerAssSavTou-JCP-18}, we introduce the following ground-state energy expression for a $N$-electron system with Hamiltonian $\hat{H}$
\begin{equation}
E^\B(\b{p})= \frac{\bra{\bar{\Psi}(\b{p})} \hat{H} \ket{\bar{\Psi}(\b{p})}}{\braket{\bar{\Psi}(\b{p})}{\bar{\Psi}(\b{p})}} + \bar{E}^\B[\rho_{\bar{\Psi}(\b{p})}],
\label{EBp}
\end{equation}
where $\bar{E}^\B[\rho]$ is the basis-set correction density functional evaluated at the density of $\bar{\Psi}(\b{p})$
\begin{equation}
\rho_{\bar{\Psi}(\b{p})}(\b{r}) = \frac{\bra{\bar{\Psi}(\b{p})} \hat{\rho}(\b{r}) \ket{\bar{\Psi}(\b{p})}}{\braket{\bar{\Psi}(\b{p})}{\bar{\Psi}(\b{p})}},
\end{equation}
where $\hat{\rho}(\b{r})$ is the density operator at point $\b{r}$. The self-consistent basis-set corrected ground-state energy is then~\cite{GinTraPraTou-JCP-21}
\begin{equation}
E_0^\B= \min_{\b{p}\in\mathbb{C}^M} E^\B(\b{p}).
\label{E0B}
\end{equation}
The role of the density functional $\bar{E}^\B[\rho]$ is to accelerate the basis convergence without altering the CBS limit. The latter point is guaranteed by imposing that $\bar{E}^\B[\rho]$ vanishes in the CBS limit, i.e. $\lim_{\B \to \text{CBS}} \bar{E}^\B[\rho] = 0$.

In practice, the minimization can be done by iteratively solving an effective Schr\"odinger equation~\cite{GinTraPraTou-JCP-21}, or, more generally, using for example the Newton method in which the current parameters are iteratively updated using the parameters changes $\bm{\Delta}\b{p}= \b{p} - \b{p}^0$ found by solving the linear equations (see, e.g., Ref.~\onlinecite{McW-BOOK-92})
\begin{equation}
\begin{pmatrix}
\b{A} & \b{B} \\
\b{B}^* & \b{A}^*
\end{pmatrix}
\begin{pmatrix}
\bm{\Delta}\b{p}\\
\bm{\Delta}\b{p}^*
\end{pmatrix}
= - 
\begin{pmatrix}
\b{g}\\
\b{g}^*
\end{pmatrix},
\label{Newton}
\end{equation}
where * designates the complex conjugate and $\b{g}$ is the energy gradient vector 
\begin{eqnarray}
g_I &=& \left. \frac{\partial E^\B(\b{p}) }{\partial p_I^*} \right|_{\b{p}=\b{p}^0}  
= \bra{\bar{\Psi}_I} \hat{H}^\B_\text{eff} \ket{\Psi_0},
\end{eqnarray}
with the effective Hamiltonian
\begin{eqnarray}
\hat{H}^\B_\text{eff}= \hat{H} + \hat{\bar{V}}^\B[\rho_{\Psi_0}],
\label{HBeff}
\end{eqnarray}
involving the basis-set correction potential operator
\begin{equation}
\hat{\bar{V}}^\B[\rho] = \int_{\mathbb{R}^3} \bar{v}^\B[\rho](\b{r}) \hat{\rho}(\b{r}) \d \b{r},
\label{VBrho}
\end{equation}
with $\bar{v}^\B[\rho](\b{r}) = \delta \bar{E}^\B[\rho]/ \delta \rho(\b{r})$. In Eq.~(\ref{Newton}), $\b{A}$ and $\b{B}$ are the energy Hessian matrices
\begin{eqnarray}
A_{I,J} &=& \left. \frac{\partial^2 E^\B(\b{p}) }{\partial p_I^* \partial p_J} \right|_{\b{p}=\b{p}^0}
\nonumber\\
&=& \bra{\bar{\Psi}_I} \hat{H}^\B_\text{eff}   -{\cal E}_0^\B\ket{\bar{\Psi}_J} + K_{I,J},
\label{AIJ}
\end{eqnarray}
where ${\cal E}_0^\B = \bra{\Psi_0} \hat{H}^\B_\text{eff} \ket{\Psi_0}$ is the energy of the effective Hamiltonian for the current wave function $\Psi_0$, and
\begin{eqnarray}
B_{I,J} &=& \left. \frac{\partial^2 E^\B(\b{p}) }{\partial p_I^* \partial p_J^*} \right|_{\b{p}=\b{p}^0}
\nonumber\\
&=& \bra{\bar{\Psi}_{I,J}} \hat{H}^\B_\text{eff} \ket{\Psi_0} + L_{I,J},
\label{BIJ}
\end{eqnarray}
involving the basis-set correction kernel contributions
\begin{eqnarray}
K_{I,J}\! = \!\!\int_{\mathbb{R}^3\times\mathbb{R}^3} \!\!\!\! \bar{f}^\B[\rho_{\Psi_0}](\b{r},\b{r}') \bra{\bar{\Psi}_I} \hat{\rho}(\b{r}) \ket{\Psi_0} \bra{\Psi_0} \hat{\rho}(\b{r}') \ket{\bar{\Psi}_J} \d\b{r} \d\b{r}', \;
\label{KIJ}
\end{eqnarray}
and
\begin{eqnarray}
L_{I,J}\! = \!\!\int_{\mathbb{R}^3\times\mathbb{R}^3} \!\!\!\! \bar{f}^\B[\rho_{\Psi_0}](\b{r},\b{r}') \bra{\bar{\Psi}_I} \hat{\rho}(\b{r}) \ket{\Psi_0} \bra{\bar{\Psi}_J} \hat{\rho}(\b{r}') \ket{\Psi_0} \d\b{r} \d\b{r}', \;
\label{LIJ}
\end{eqnarray}
with $\bar{f}^\B[\rho](\b{r},\b{r}') = \delta^2 \bar{E}^\B[\rho]/ \delta \rho(\b{r}) \delta \rho(\b{r'})$. 

At the end of the optimization, provided that we have reached the global energy minimum, the current parameters $\b{p}^0$ are the optimal ground-state parameters. To make clear the link with our previous work~\cite{GinTraPraTou-JCP-21}, we note that the present energy minimization is equivalent to solving the following effective Schr\"odinger equation projected in the basis of the first-order wave-function derivatives $\{ \ket{\bar{\Psi}_I} \}_{I=1,...,M}$
\begin{eqnarray}
\bra{\bar{\Psi}_I}\hat{H}_\text{eff}^\B -{\cal E}_0^\B \ket{\Psi_0} = 0.
\label{Schroeq}
\end{eqnarray}
Since $\braket{\bar{\Psi}_I}{\Psi_0}=0$, Eq.~(\ref{Schroeq}) is indeed equivalent to having a zero energy gradient, i.e. $g_I=\bra{\bar{\Psi}_I} \hat{H}^\B_\text{eff} \ket{\Psi_0}=0$.

\subsection{General linear-response equations}

Starting from the optimal ground state, we now add a time-dependent perturbation operator $\hat{V}(t)$ to the Hamiltonian,
\begin{eqnarray}
\hat{H}(t) = \hat{H} + \hat{V}(t),
\end{eqnarray}
where $\hat{V}(t)$ is chosen as a periodic monochromatic electric-dipole interaction of frequency $\omega$
\begin{eqnarray}
\hat{V}(t) = - \hat{\b{d}} \cdot \bm{\epsilon}^+ e^{-\i \omega t} - \hat{\b{d}} \cdot \bm{\epsilon}^- e^{+\i \omega t},
\end{eqnarray}
where $\hat{\b{d}} = - \int_{\mathbb{R}^3} \b{r} \; \hat{\rho}(\b{r}) \d\b{r}$ is the dipole-moment operator, and $\bm{\epsilon}^+$ and $\bm{\epsilon}^-$ are the electric-field strengths for the positive and negative frequency terms (taken as different for intermediate derivations but ultimately we must have $\bm{\epsilon}^+=\bm{\epsilon}^-$).

The wave function $\ket{\bar{\Psi}(\b{p}(t))}$ will now depend on time through the parameters $\b{p}(t)=\b{p}^0 + \bm{\Delta}\b{p}(t)$ where $\b{p}^0$ are the optimal ground-state parameters and $\bm{\Delta}\b{p}(t)$ are the time variations of the parameters which can be written as
\begin{eqnarray}
\bm{\Delta}\b{p}(t) = \b{p}^+ e^{-\i \omega t} + \b{p}^- e^{+\i \omega t},
\end{eqnarray}
where $\b{p}^+ \in \mathbb{C}^M$ and $\b{p}^- \in \mathbb{C}^M$ are the Fourier components. The ground-state energy expression in Eq.~(\ref{EBp}) is replaced by the quasi-energy expression~\cite{ChrJorHat-IJQC-98,SalVahHelAgr-JCP-02,HelCorJorKriOlsRuu-CR-12,FroKneJen-JCP-13,NorRuuSau-BOOK-18,Tou-LEC-18}
\begin{eqnarray}
Q^\B(\bm{\epsilon}^+,\bm{\epsilon}^-;\b{p}^+,\b{p}^-) = \;\;\;\;\;\;\;\;\;\;\;\;\;\;\;\;\;\;\;\;\;\;\;\;\;\;\;\;\;\;\;\;\;\;\;\;\;\;\;\;\;\;\;\;\;\;\;\;
\nonumber\\ \frac{1}{T} \int_0^T \Biggl\{ \frac{\bra{\bar{\Psi}(\b{p}(t))} \hat{H}(t) -i \frac{\partial}{\partial t} \ket{\bar{\Psi}(\b{p}(t))}}{\braket{\bar{\Psi}(\b{p}(t))}{\bar{\Psi}(\b{p}(t))}} 
 + \bar{E}^\B[\rho_{\bar{\Psi}(\b{p}(t))}] \Biggl\} \; \d t,
\label{QB}
\end{eqnarray}
where $T= 2\pi/\omega$ is the period. 
Note that, in the definition of the quasi-energy in Eq.~(\ref{QB}), the same basis-set correction density functional $\bar{E}^\B[\rho]$ used for the ground-state calculation is employed, which is known as the adiabatic approximation. This approximation is almost always used in time-dependent DFT calculations of excitation energies (see, e.g, Refs.~\cite{Cas-INC-96,BauAhl-CPL-96}). Due to this approximation, the basis-set correction contribution to the quasi-energy is a local functional of time. Overcoming this approximation would require the complicated task of developing a quasi-energy basis-set correction contribution having the form a non-local functional of time, i.e. depending on all the time history. We do not attempt to do that in the present work.
The optimal quasi-energy $Q_0^\B(\bm{\epsilon}^+,\bm{\epsilon}^-)$ is a stationary value of $Q^\B(\bm{\epsilon}^+,\bm{\epsilon}^-;\b{p}^+,\b{p}^-)$ with respect to variations of the parameters $\b{p}^+$ and $\b{p}^-$, which we write as
\begin{eqnarray}
Q_0^\B(\bm{\epsilon}^+,\bm{\epsilon}^-) \in \stat{(\b{p}^+,\b{p}^-) \in \mathbb{C}^{2M}} Q^\B(\bm{\epsilon}^+,\bm{\epsilon}^-;\b{p}^+,\b{p}^-),
\end{eqnarray}
where ``stat'' refers to the set of stationary values.
In the zero electric-field limit ($\bm{\epsilon}^+=\bm{\epsilon}^-=\b{0}$), vanishing parameters $\b{p}^+=\b{p}^-= \b{0}$ are optimal and the corresponding optimal quasi-energy reduces to the ground-state energy, i.e. $Q_0^\B(\b{0},\b{0})=E_0^\B$.

The optimal quasi-energy allows one to define the dynamic dipole polarizability tensor as 
\begin{eqnarray}
\alpha_{i,j}^\B(\omega) = - \left. \frac{\partial^2 Q_0^\B(\bm{\epsilon}^+,\bm{\epsilon}^-)}{\partial \epsilon_i^- \partial \epsilon_j^+} \right|_{\bm{\epsilon}^\pm=\b{0}},
\end{eqnarray}
where $i$ and $j$ refer to 3D Cartesian components. Calculating this second-order derivative using the chain rule via the optimal parameters $\b{p}^+$ and $\b{p}^-$ (which implicitly depend on $\bm{\epsilon}^+$ and $\bm{\epsilon}^-$) leads to (see, e.g., Refs.~\onlinecite{NorRuuSau-BOOK-18,Tou-LEC-18})
\begin{eqnarray}
\alpha_{i,j}^\B(\omega) = 
\begin{pmatrix}
\b{V}_i \\
\b{V}_i^*
\end{pmatrix}^\dagger
\begin{pmatrix}
\bm{\Lambda}(\omega) & \bm{\Xi} \\
\bm{\Xi}^* & \bm{\Lambda}(-\omega)^*
\end{pmatrix}^{-1} 
\begin{pmatrix}
\b{V}_j \\
\b{V}_j^*
\end{pmatrix},
\label{alphaijmatrix}
\end{eqnarray}
where $\bm{\Lambda}(\omega)$ and $\bm{\Xi}$ are the quasi-energy Hessian matrices
\begin{eqnarray}
\Lambda_{I,J}(\omega) &=& \left. \frac{\partial^2 Q^\B(\bm{\epsilon}^+,\bm{\epsilon}^-;\b{p}^+,\b{p}^-) }{\partial p_I^{+*} \partial p_J^+} \right|_{\substack{\b{p}^\pm=\b{0} \\ \bm{\epsilon}^\pm=\b{0}}}
= A_{I,J} - \omega S_{I,J}, \;
\end{eqnarray}
and
\begin{eqnarray}
\Xi_{I,J} &=& \left. \frac{\partial^2 Q^\B(\bm{\epsilon}^+,\bm{\epsilon}^-;\b{p}^+,\b{p}^-) }{\partial p_I^{+*} \partial p_J^{-*}} \right|_{\substack{\b{p}^\pm=\b{0} \\ \bm{\epsilon}^\pm=\b{0}}}
= B_{I,J}, 
\end{eqnarray}
where $A_{I,J}$ and $B_{I,J}$ are given in Eqs.~(\ref{AIJ}) and (\ref{BIJ}), and $S_{I,J}$ is the overlap matrix of the first-order wave-function derivatives
\begin{eqnarray}
S_{I,J} = \braket{\bar{\Psi}_I}{\bar{\Psi}_J}.
\end{eqnarray}
Equation~(\ref{alphaijmatrix}) also involves the perturbed energy gradient vector 
\begin{eqnarray}
V_{j,I} = \left. \frac{\partial^2 Q^\B(\bm{\epsilon}^+,\bm{\epsilon}^-;\b{p}^+,\b{p}^-) }{\partial \epsilon_j^+ \partial p_I^{+*}} \right|_{\substack{\b{p}^\pm=\b{0} \\ \bm{\epsilon}^\pm=\b{0}}} 
= - \bra{\bar{\Psi}_I} \hat{d}_j \ket{\Psi_0},
\end{eqnarray}
corresponding to transition dipole-moment matrix elements.

Finally, the poles in $\omega$ of the dynamic dipole polarizability $\alpha_{i,j}^\B(\omega)$ provide $M$ positive excitation energies $\{\omega_n^\B\}$ (and $M$ opposite deexcitation energies), which can be found from the following generalized eigenvalue equation (see, e.g., Ref.~\onlinecite{McW-BOOK-92})
\begin{equation}
\begin{pmatrix}
\b{A} & \b{B} \\
\b{B}^* & \b{A}^*
\end{pmatrix}
\begin{pmatrix}
\b{X}_n\\
\b{Y}_n
\end{pmatrix}
= \omega_n^\B 
\begin{pmatrix}
\b{S} & \b{0} \\
\b{0} & -\b{S}
\end{pmatrix}
\begin{pmatrix}
\b{X}_n\\
\b{Y}_n
\end{pmatrix},
\label{ABBASS}
\end{equation}
where $(\b{X}_n,\b{Y}_n)$ are eigenvectors. The obtained excitation energies $\{\omega_n^\B\}$ include the basis-set correction through the potential $\bar{v}^\B[\rho](\b{r})$ in Eq.~(\ref{VBrho}) and kernel $\bar{f}^\B[\rho](\b{r},\b{r}')$ in Eqs.~(\ref{KIJ}) and~(\ref{LIJ}), and may be expected to converge faster to their CBS limit, provided good enough approximations are used for the basis-set correction functional used for $\bar{v}^\B[\rho](\b{r})$ and $\bar{f}^\B[\rho](\b{r},\b{r}')$. Obviously, the corresponding basis-set corrected total energy of the n$^\text{th}$ excited state is given by
\begin{equation}
E_n^\B = E_0^\B + \omega_n^\B,
\label{EnB}
\end{equation}
and could also be expected to converge faster to its CBS limit, if the basis-set correction functional is good enough.

\subsection{Linear-response equations for configuration-interaction wave functions}

We now give the more specific form of the linear-response equations for configuration-interaction (CI) wave functions. Given a set of $M$ orthonormal configurations $\{\ket{\Phi_I}\}$, the CI wave function is parametrized as 
\begin{equation}
\ket{\Psi(\b{p})} = \sum_{I=1}^M p_I \ket{\Phi_I}.
\label{}
\end{equation}
The ground-state parameters are assumed to be real valued and are denoted by $p_I^0 = c_I$, so that the ground-state wave function is $\ket{\Psi_0}=\sum_{I=1}^M c_I \ket{\Phi_I}$. From Eq.~(\ref{PsiIandPsiIJ}), the first-order and second-order derivatives of the intermediately normalized wave function are found to be
\begin{equation}
\ket{\bar{\Psi}_I} = \ket{\Phi_I} - c_I \; \ket{\Psi_0},
\label{}
\end{equation}
and
\begin{equation}
\ket{\bar{\Psi}_{I,J}} = 2 c_I c_J \; \ket{\Psi_0} - c_J \; \ket{\Phi_I} - c_I \; \ket{\Phi_J}.
\label{}
\end{equation}
In a spin-restricted formalism with a set of real-valued orthonormal orbitals $\{ \varphi_i \} \subset {\calh^\B}$, the linear-response matrices in Eq.~(\ref{ABBASS}) now become
\begin{eqnarray}
A_{I,J} &=& \bra{\Phi_I} \hat{H}^\B_\text{eff}  -{\cal E}_0^\B\ket{\Phi_J} + K_{I,J},
\label{}
\end{eqnarray}
\begin{eqnarray}
B_{I,J} &=& K_{I,J},
\label{}
\end{eqnarray}
\begin{eqnarray}
S_{I,J} &=& \delta_{I,J} - c_I c_J,
\label{}
\end{eqnarray}
where the kernel contribution takes the form
\begin{eqnarray}
K_{I,J} &=& \sum_{i,j,k,l} \Delta \gamma_{i,j}^I \; \Delta \gamma_{k,l}^J \; \bar{f}^\B_{i,j,k,l},
\label{KIJijkl}
\end{eqnarray}
with $i,j,k,l$ referring to spatial orbitals. In Eq.~(\ref{KIJijkl}), we have introduced
\begin{eqnarray}
\Delta \gamma_{i,j}^I = \gamma_{i,j}^I - c_I \gamma_{i,j},
\label{}
\end{eqnarray}
where $\gamma_{i,j}$ and $\gamma_{i,j}^I$ are the ground-state and transition density matrices, respectively,
\begin{eqnarray}
\gamma_{i,j} = \bra{\Psi_0} \hat{E}_{i,j} \ket{\Psi_0} = \sum_{I=1}^M \sum_{J=1}^M c_I c_J \bra{\Phi_I} \hat{E}_{i,j} \ket{\Phi_J},
\label{}
\end{eqnarray}
\begin{eqnarray}
\gamma_{i,j}^I = \bra{\Phi_I}\hat{E}_{i,j} \ket{\Psi_0} = \sum_{J=1}^M c_J \bra{\Phi_I}\hat{E}_{i,j} \ket{\Phi_J},
\label{}
\end{eqnarray}
where $\hat{E}_{i,j} = \hat{a}_{i\uparrow}^\dagger \hat{a}_{j\uparrow} + \hat{a}_{i\downarrow}^\dagger \hat{a}_{j\downarrow}$ is the spin-summed one-electron density-matrix operator in second quantization. Finally, in Eq.~(\ref{KIJijkl}), $\bar{f}^\B_{i,j,k,l}$ are the matrix elements of the basis-set correction kernel $\bar{f}^\B[\rho_{\Psi_0}](\b{r},\b{r}')$ over the spatial orbitals
\begin{eqnarray}
\bar{f}^\B_{i,j,k,l} = \int_{\mathbb{R}^3\times\mathbb{R}^3} \bar{f}^\B[\rho_{\Psi_0}] (\b{r},\b{r}') \; \varphi_i(\b{r}) \varphi_j(\b{r}) \varphi_k(\b{r}') \varphi_l(\b{r}') \d\b{r} \d\b{r}'. \;
\label{}
\end{eqnarray}

\section{One-dimensional model system} 
\label{sec:1Dmodel}

\subsection{Description of the model and exact solutions}

\begin{table}[]
\caption{Exact total energies of the first 5 eigenstates of even-parity and singlet symmetries for the 1D two-electron Hooke-type atom with $\omega_0 = 1$, and corresponding excitation energies. All energies are in hartree.}
\label{tab:exact}
\begin{tabular}{lcc}
\hline\hline
State $(n,m)$ $\;\;\;$  & $\;\;\;$ Total energy $E_{n,m}$ & $\;\;\;$ Excitation energy $E_{n,m}-E_{0,0}$ \\
\hline
$(0,0)$ & 1.306746 & \\
$(0,2)$ & 3.187051 & 1.880305 \\
$(2,0)$ & 3.306746 & 2.000000\\
$(0,4)$ & 5.144734 & 3.837988 \\
$(2,2)$ & 5.187051 & 3.880305 \\
\hline\hline
\end{tabular}
\end{table}

We consider the 1D two-electron Hooke-type atom studied in Refs.~\onlinecite{MagBur-PRA-04,MaiZhaCavBur-JCP-04,TraGinTou-JCP-22}. We work first in the infinite-dimensional spin-free one-electron Hilbert space $\calh = L^2(\mathbb{R},\mathbb{C})$ and the associated non-antisymmetrized tensor-product two-electron Hilbert space ${\cal H} = \calh \otimes \calh$. The Hamiltonian is
\begin{eqnarray}
\hat{H} &=& -\frac{1}{2} \frac{\partial^2}{\partial x_1^2} - \frac{1}{2} \frac{\partial^2}{\partial x_{2}^2} + \frac{1}{2} \omega_0^2 x_1^2 + \frac{1}{2} \omega_0^2 x_2^2 + \delta(x_1-x_2), \;\;\;\;\;\;\;
\label{Eq:Hamiltonian}
\end{eqnarray}
involving a harmonic external potential of curvature $\omega_0^2$ (which will be chosen to $1$ throughout this study) and a Dirac-delta two-electron interaction. The latter two-electron interaction generates in 1D the same s-wave electron-electron cusp as the Coulomb interaction does in 3D, and it is thus an appropriate model to study the basis convergence~\cite{TraGinTou-JCP-22}. This 1D two-electron Hooke-type atom can be considered as the 1D analog of the well-known 3D two-electron Hooke atom (see, e.g., Refs.~\onlinecite{Tau-PRA-93,Kin-TCA-96}).

Expressed with the center-of-mass (cm) coordinate $X=(x_1+x_2)/2$ and the relative (rel) coordinate $x_{12}=x_1-x_2$, the Hamiltonian is separable~\cite{TraGinTou-JCP-22} and its eigenvalues are
\begin{eqnarray}
E_{n,m} = E^\text{cm}_{n} + E^\text{rel}_{m} \; \text{for} \; n\in \mathbb{N}\; \text{and}\; m\in \mathbb{N},
\end{eqnarray}
where $E^\text{cm}_{n} = \omega_0 (n + 1/2)$ is the center-of-mass contribution and $E^\text{rel}_{m}$ is the relative contribution, 
\begin{eqnarray}
E^\text{rel}_{m} = 
\begin{cases}
\omega_0 (\nu_{m} + 1/2), \; \text{for}\; m \;\text{even},\\[0.2cm]
\omega_0 (m + 1/2), \; \text{for}\;  m \; \text{odd},\\
\end{cases}
\end{eqnarray}
with the real numbers $\nu_{m}$ being the solutions of the equation~\cite{BusHuy-JPB-03,MagBur-PRA-04}
\begin{eqnarray}
2\sqrt{2\omega_0} \; \frac{\Gamma\left( -\frac{\nu_m}{2} + \frac{1}{2}\right)}{\Gamma\left( -\frac{\nu_m}{2} \right)} = -1,
\end{eqnarray}
where $\Gamma$ is the gamma function. The associated eigenfunctions are
\begin{eqnarray}
\Psi_{n,m}(X,x_{12}) = \psi_n^\text{cm}(X) \psi_m^\text{rel}(x_{12}),
\end{eqnarray}
where the center-of-mass eigenfunctions are $\psi_n^\text{cm}(X) = f_n^{2\omega_0}(X)$ and the relative eigenfunctions are
\begin{eqnarray}
\psi^\text{rel}_{m}(x_{12}) = 
\begin{cases}
c_m D_{\nu_m}(\sqrt{\omega_0}|x_{12}|), \; \text{for}\; m \;\text{even},\\[0.2cm]
f_m^{\omega_0/2}(x_{12}), \; \text{for}\;  m \; \text{odd},\\
\end{cases}
\end{eqnarray}
where $D_{\nu_m}$ is the parabolic cylinder function~\cite{AbrSte-BOOK-83} and $c_m$ is a normalization constant. Here, $f_n^{\omega_0}$ designates the Hermite functions (i.e., quantum harmonic-oscillator eigenfunctions) for the frequency $\omega_0$
\begin{eqnarray}
f_n^{\omega_0}(x) = N^{\omega_0}_n \; H_n(\sqrt{\omega_0}x) \; e^{-\omega_0 x^2/2},
\end{eqnarray}
where $H_n$ are the Hermite polynomials, $N_n^{\omega_0} = (2^n n!)^{-1/2} (\omega_0/\pi)^{1/4}$ is the normalization factor. As announced, for even $m$, the relative eigenfunctions has the familiar s-wave cusp condition: $\psi_m^\text{rel}(x_{12}) = \psi_m^\text{rel}(0)[1+(1/2)|x_{12}| + O(x_{12}^2)]$ (see Ref.~\onlinecite{TraGinTou-JCP-22}).

We will only consider eigenstates of even-parity symmetry [i.e., invariant under the transformation $(x_1,x_2) \to (-x_1,-x_2)$] and singlet symmetry (i.e., invariant under the exchange $x_1 \leftrightarrow x_2$), corresponding to the eigenstates with both even quantum numbers $n$ and $m$. The exact total energies of the first 5 of these eigenstates, as well as the corresponding excitation energies, are given in Table~\ref{tab:exact} for $\omega_0 = 1$.

\begin{figure}
\centering
\includegraphics[scale=0.3,angle=-90]{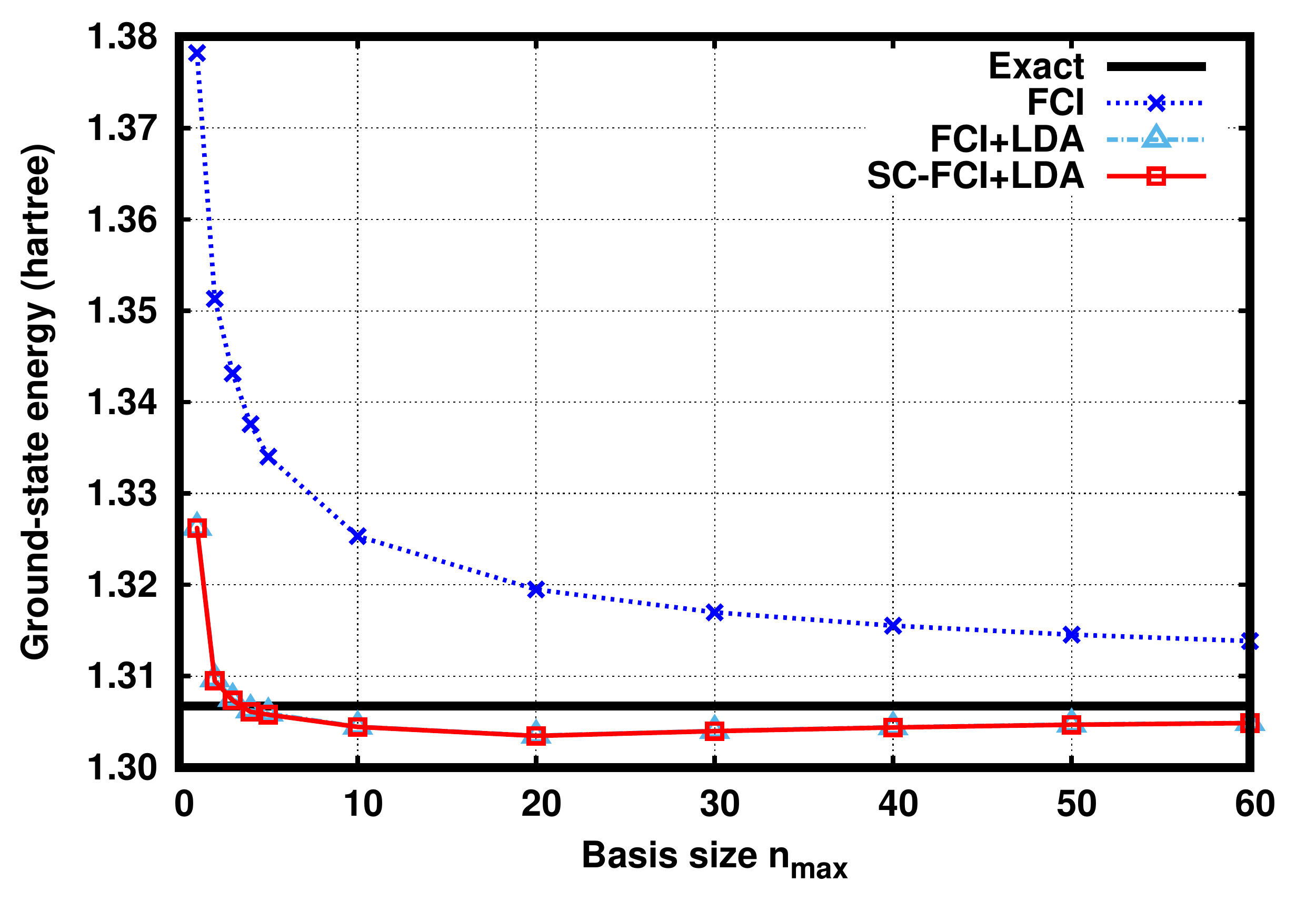}
\caption{Ground-state energy of the 1D two-electron Hooke-type atom with $\omega_0 = 1$ calculated by the standard FCI method, the non-self-consistent FCI+LDA and self-consistent SC-FCI+LDA basis-set corrected methods as a function of the basis size $n_\text{max}$.
}
\label{Fig:GroundState}
\end{figure}

\subsection{Full-configuration-interaction calculation in a basis set}

We consider finite basis sets of Hermite (or Hermite-Gauss) functions 
\begin{equation}
\B = \{ f_n^{\omega_0} \}_{n=1,...,n_\text{max}},
\end{equation}
with a fixed parameter $\omega_0 = 1$ and a variable maximal quantum number $n_\text{max}$ determining the basis size. The one-electron and two-electron Hilbert spaces corresponding to this basis set are ${\calh}^\B = \text{span}(\B)$ and ${\cal H}^\B = {\calh}^\B \otimes {\calh}^\B$. 

\begin{figure*}
\centering
\includegraphics[scale=0.3,angle=-90]{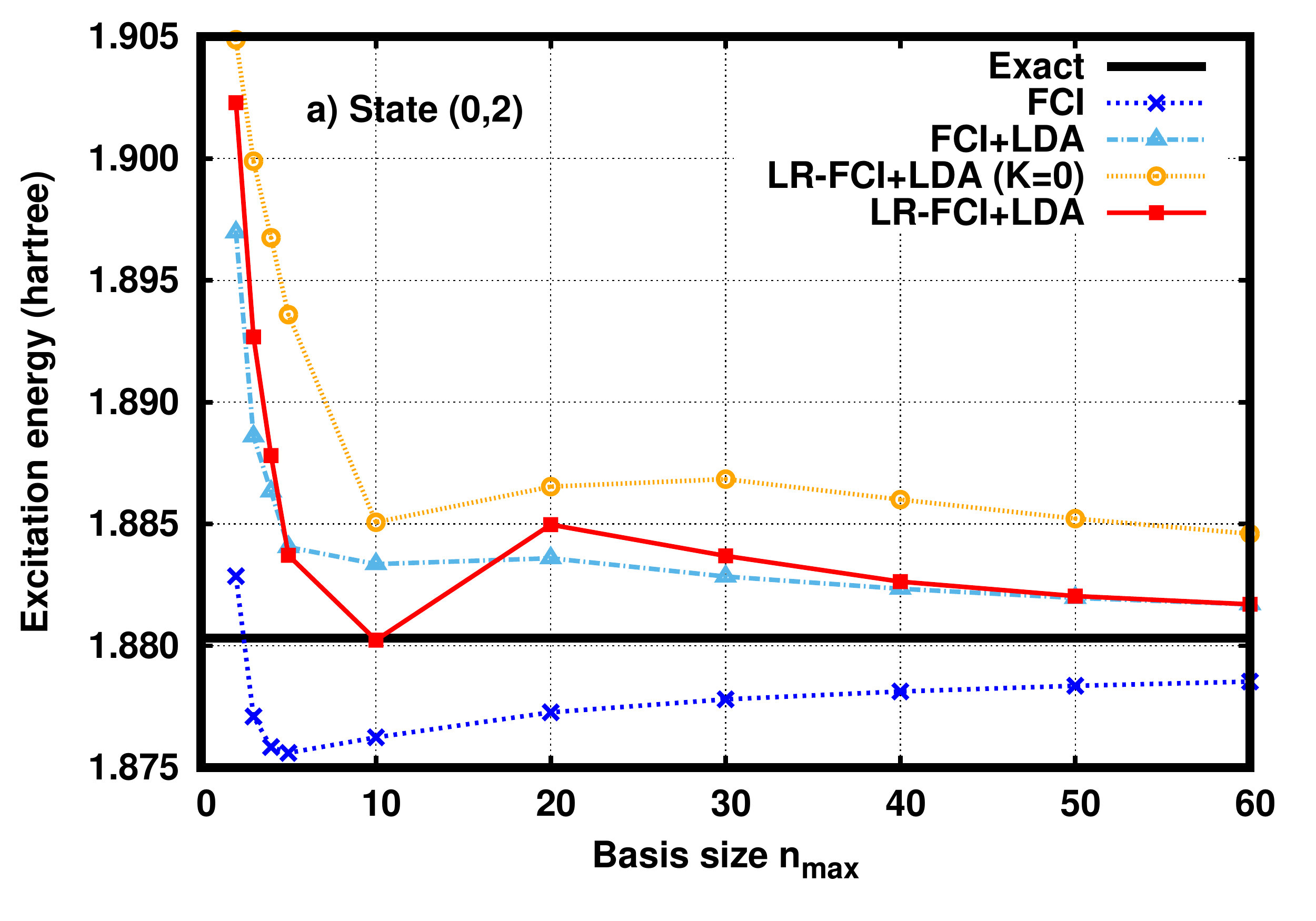}
\includegraphics[scale=0.3,angle=-90]{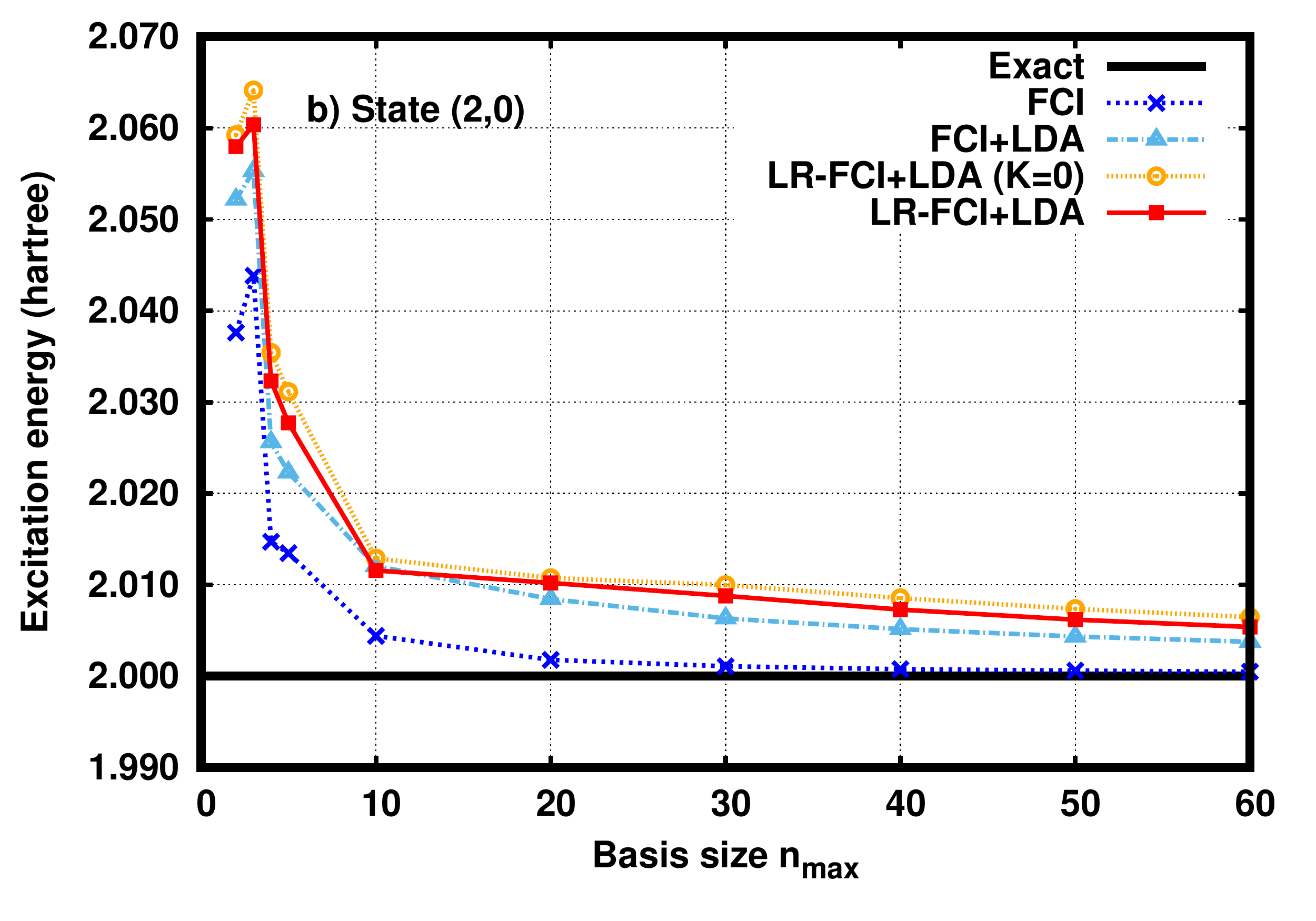}
\includegraphics[scale=0.3,angle=-90]{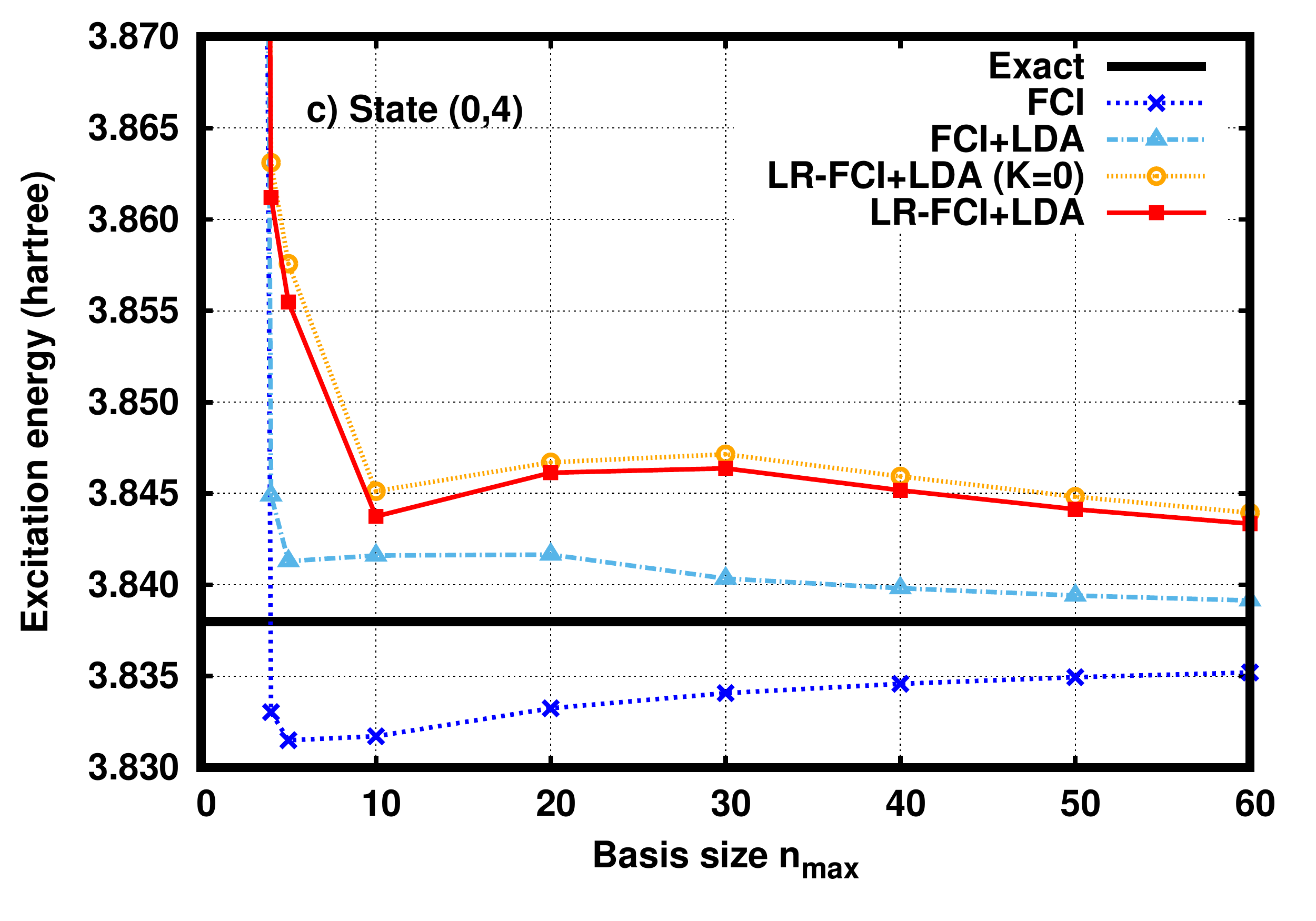}
\includegraphics[scale=0.3,angle=-90]{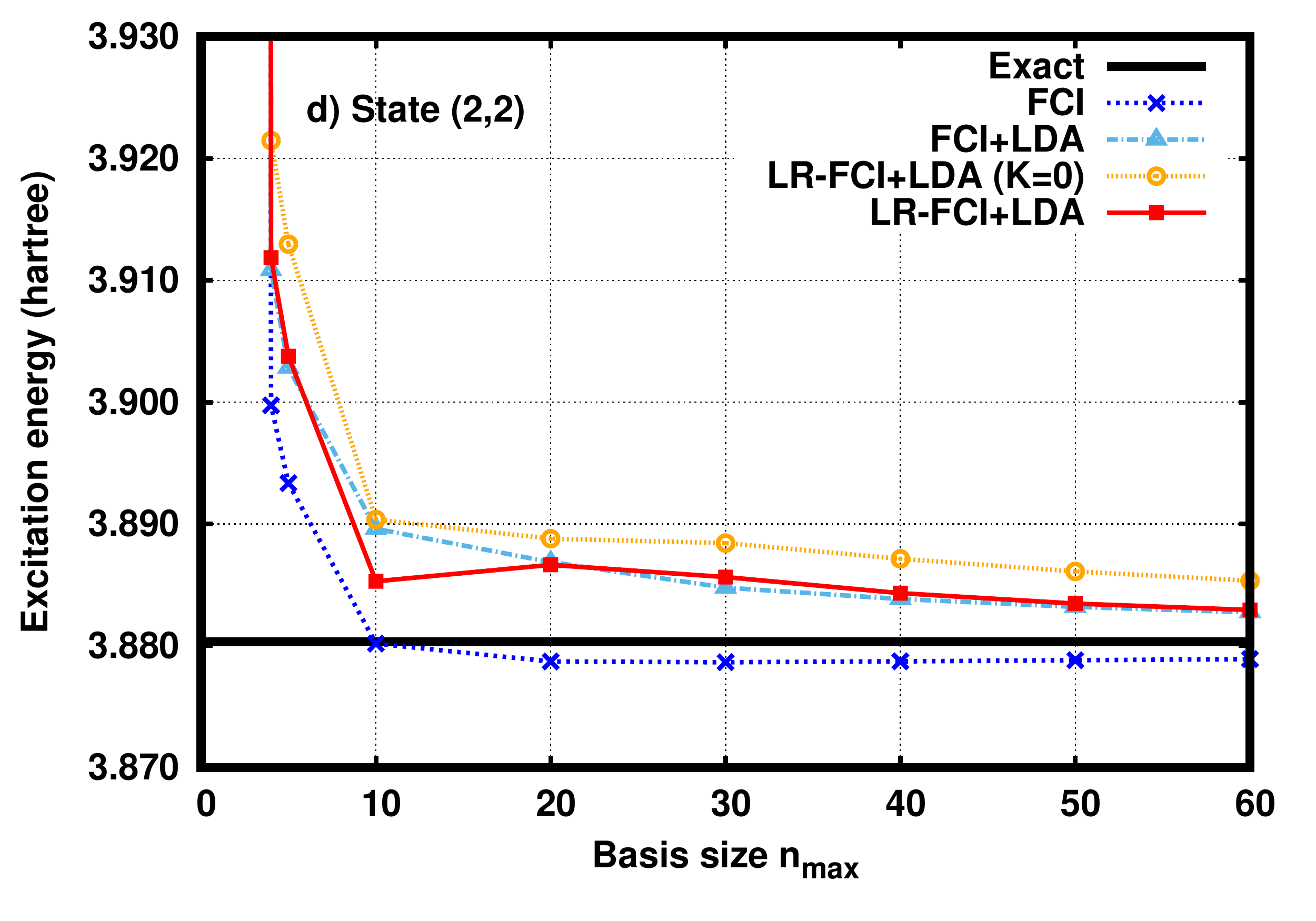}
\caption{Excitation energies of the states $(0,2)$, $(2,0)$, $(0,4)$, and $(2,2)$ of the 1D two-electron Hooke-type atom with $\omega_0 = 1$ calculated by the standard FCI method and the LR-FCI+LDA basis-set corrected method as a function of the basis size $n_\text{max}$. For comparison, excitation energies obtained with a zero basis-set correction kernel [LR-FCI+LDA (K=0)] and with a non-self-consistent state-specific approach (FCI+LDA) are also shown.}
\label{Fig:ExcitedStates}
\end{figure*}

For several values of $n_\text{max}$, we first perform a Hartree-Fock (HF) calculation to obtain the set of orthonormal HF orbitals $\{ \varphi_i \}$, and we then perform a full-configuration-interaction (FCI) calculation for the states of even-parity symmetry. The parameterized FCI wave function is thus $\ket{\Psi_\FCI(\b{p})} = \sum_{I=1}^M p_I \ket{\Phi_I}$ where $\ket{\Phi_I} = \ket{\varphi_{I_1}} \otimes \ket{\varphi_{I_2}}$, and the orbitals $\varphi_{I_1}$ and $\varphi_{I_2}$ are restricted to be of the same parity symmetry. In Fig.~\ref{Fig:GroundState} we report the FCI ground-state energy $E_{0,\FCI} = \bra{\Psi_{0,\FCI}} \hat{H} \ket{\Psi_{0,\FCI}}$, where $\Psi_{0,\FCI}$ is the FCI ground-state wave function, as a function of the basis size $n_\text{max}$. As expected, the FCI ground-state energy slowly converges toward the exact ground-state energy as $n_\text{max}$ increases. The convergence rate is compatible with the theoretical convergence rate of $1/n_\text{max}^{1/2}$ determined in Ref.~\onlinecite{TraGinTou-JCP-22}.

\begin{figure*}
\centering
\includegraphics[scale=0.3,angle=-90]{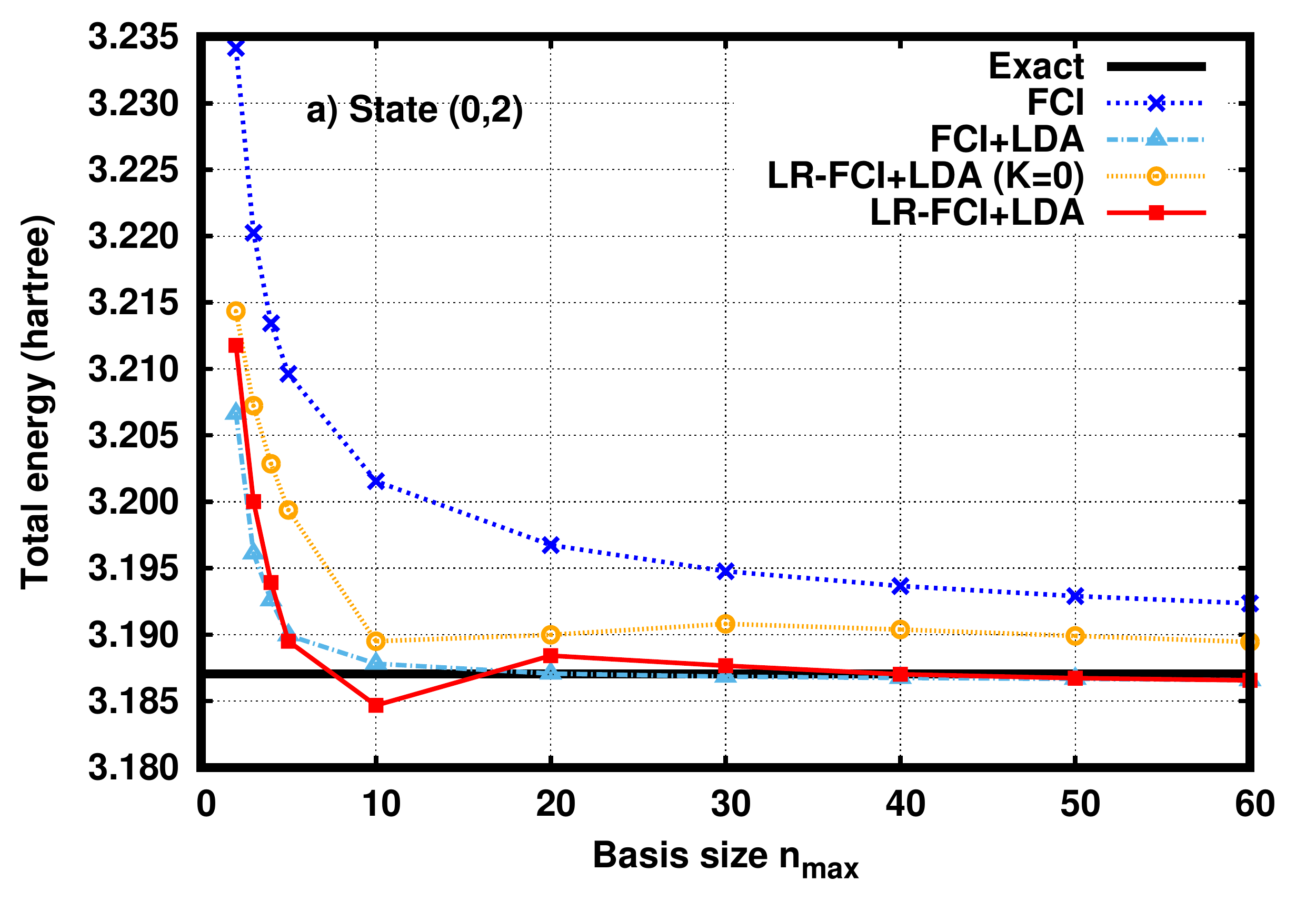}
\includegraphics[scale=0.3,angle=-90]{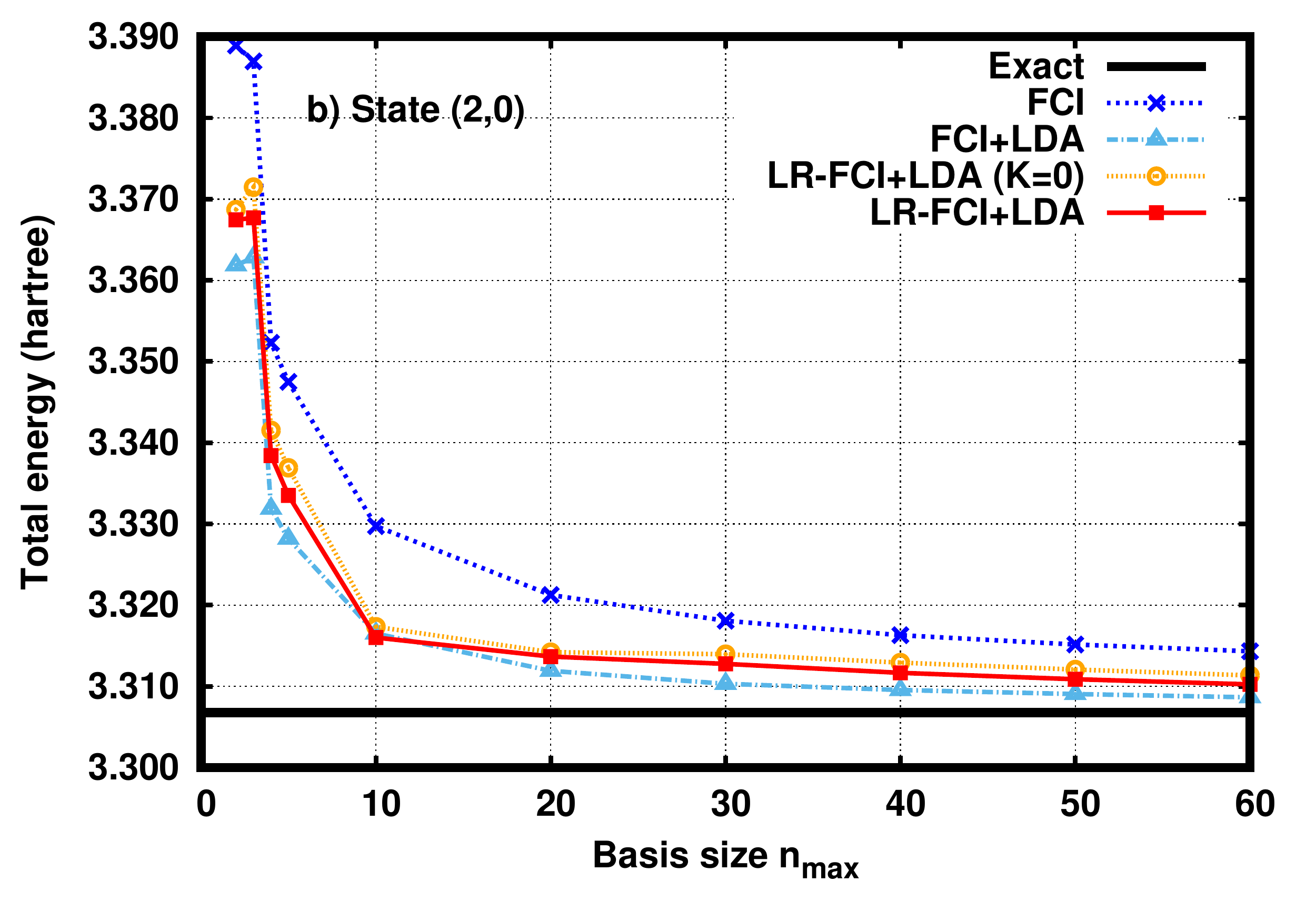}
\includegraphics[scale=0.3,angle=-90]{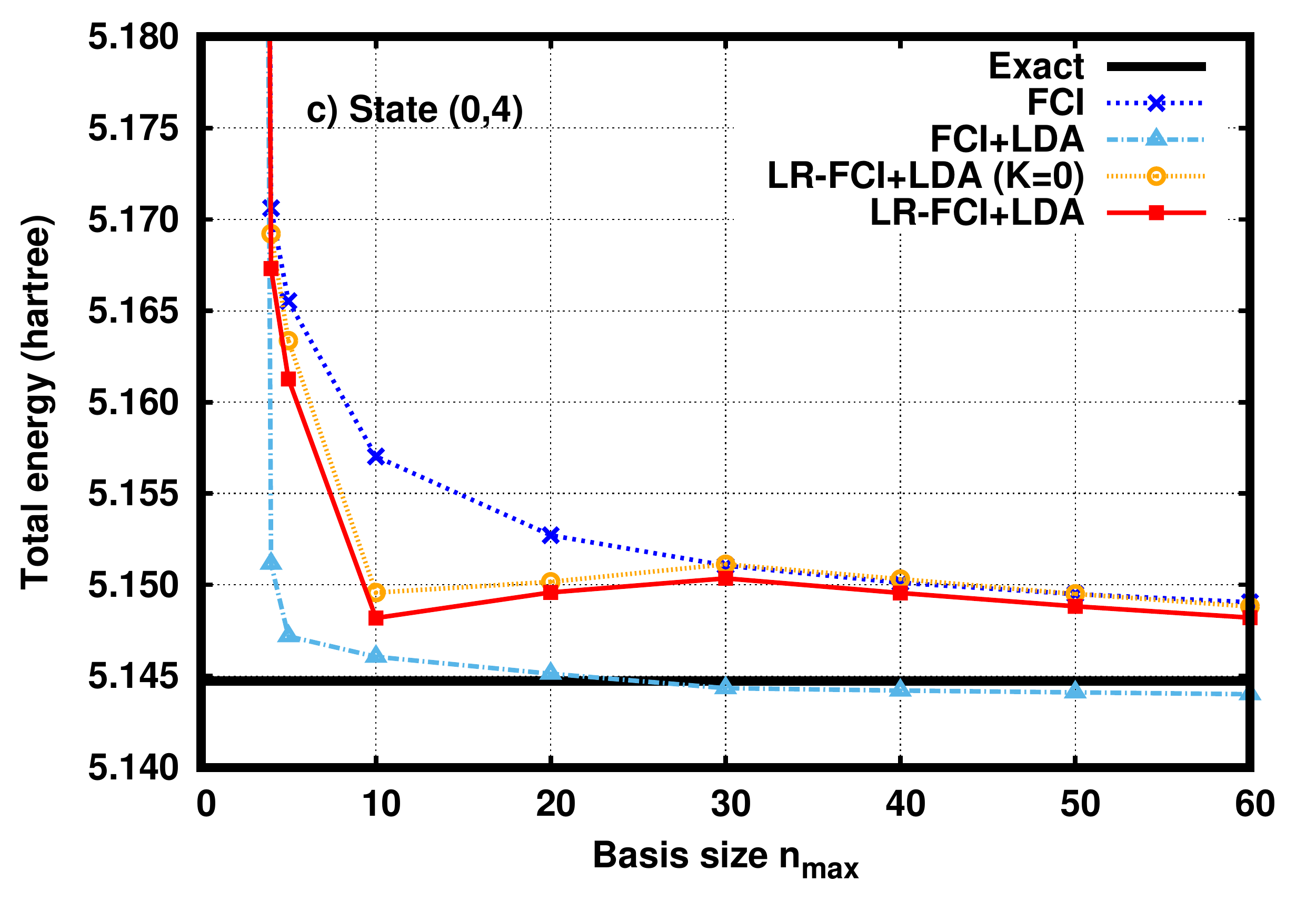}
\includegraphics[scale=0.3,angle=-90]{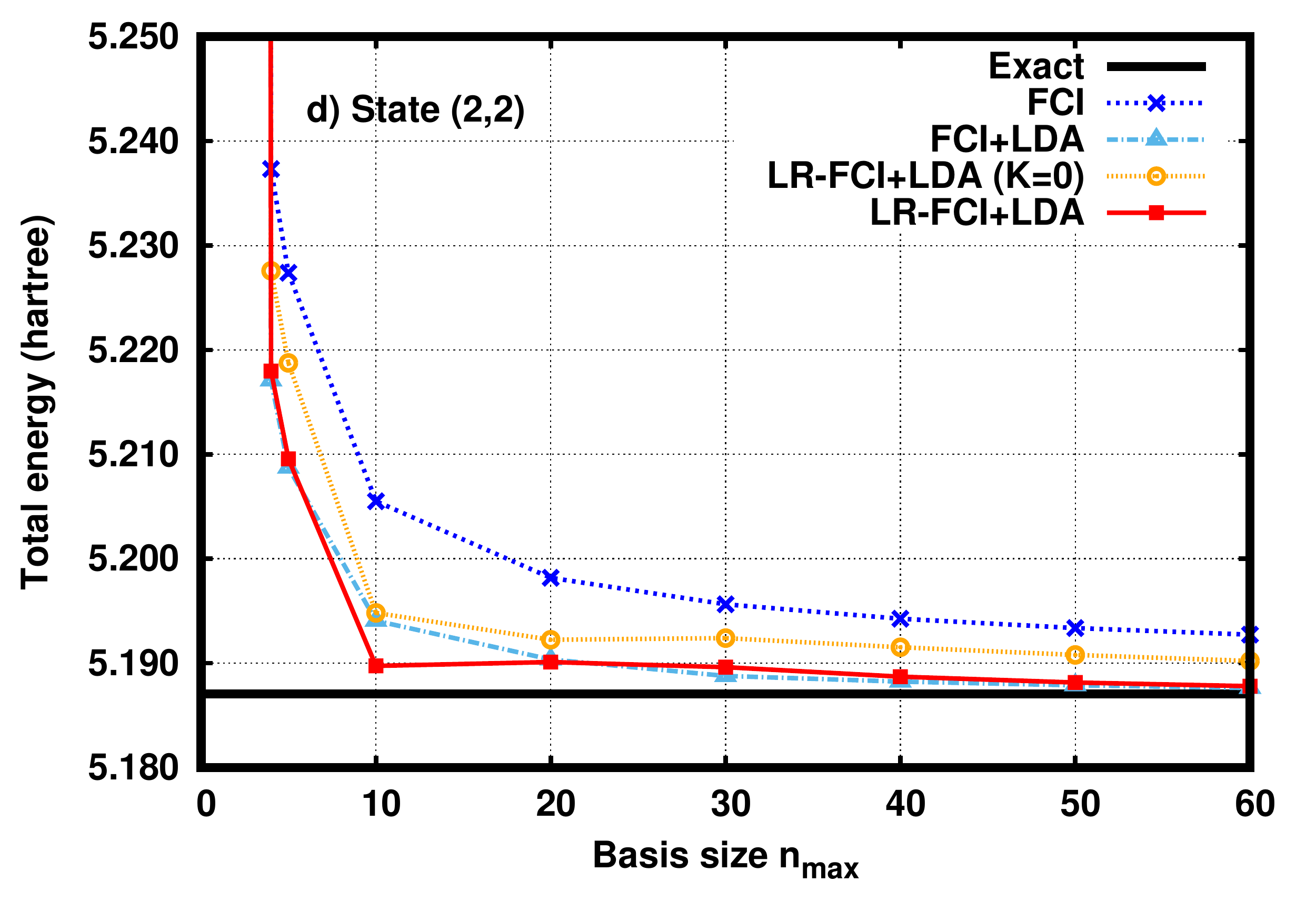}
\caption{Excited-state total energies of the states $(0,2)$, $(2,0)$, $(0,4)$, and $(2,2)$ of the 1D two-electron Hooke-type atom with $\omega_0 = 1$ calculated by the standard FCI method and the LR-FCI+LDA basis-set corrected method as a function of the basis size $n_\text{max}$. For comparison, excited-state total energies obtained with a zero basis-set correction kernel [LR-FCI+LDA (K=0)] and with a non-self-consistent state-specific approach (FCI+LDA) are also shown.}
\label{Fig:ExcitedStatesTotalEnergy}
\end{figure*}

We construct a local-density approximation (LDA) for the basis-set correction functional $\bar{E}^{\B}[\rho]$
\begin{equation}
\bar{E}^{\B}_\LDA[\rho] = \int_{\mathbb{R}} \, \rho(\b{r}) \bar{\varepsilon}^{\mathcal{B}}(\rho(\b{r})) \, \d\b{r},
\label{EBLDA}
\end{equation}
where the energy per particle $\bar{\varepsilon}^{\mathcal{B}}(\rho)$ is defined in exactly the same way as in Ref.~\onlinecite{TraGinTou-JCP-22}, i.e. as the complementary multideterminant correlation energy per particle of a two-electron finite uniform electron gas with electron-electron interaction projected in the basis set $\B$. For convenience, we fit the numerically calculated energy per particle $\bar{\varepsilon}^{\mathcal{B}}(\rho)$ to a rational fraction
\begin{equation}
\bar{\varepsilon}^{\mathcal{B}}(\rho) \approx \frac{\sum_{i=0}^4 a^{\mathcal{B}}_i \rho^i}{1 + \sum_{j=1}^4 b^{\mathcal{B}}_j \rho^j},
\label{epsilonB}
\end{equation}
where the values of the coefficients $a_i^{\mathcal{B}}$ and $b^{\mathcal{B}}_j$ for each basis size $n_\text{max}$ are given in the Supplementary Information.

We perform a ground-state FCI calculation including self-consistently the basis-set correction LDA functional according to Eq.~(\ref{E0B}). The required LDA basis-set correction potential is obtained by straightforward differentiation of Eq.~(\ref{EBLDA})
\begin{equation}
\bar{v}^\B_\LDA[\rho](\b{r}) = \bar{\varepsilon}^{\mathcal{B}}(\rho(\b{r})) + \rho(\b{r}) \left. \frac{\d \bar{\varepsilon}^{\mathcal{B}}(\rho)}{\d \rho} \right|_{\rho = \rho(\b{r})}.
\label{vBLDAr}
\end{equation}
The resulting energy, labelled as SC-FCI+LDA, is reported as a function of the basis size $n_\text{max}$ in Fig.~\ref{Fig:GroundState}. We see that the basis-set correction LDA functional is very effective in reducing the basis-set incompleteness error, resulting in a fast convergence of the SC-FCI+LDA ground-state energy toward the exact ground-state energy. For comparison, we also show in Fig.~\ref{Fig:GroundState} the non-self-consistent approximation~\cite{GinPraFerAssSavTou-JCP-18,TraGinTou-JCP-22}, labelled as FCI+LDA,
\begin{equation}
E_{0,\text{FCI+LDA}} = E_{0,\FCI} + \bar{E}^{\B}_\LDA[\rho_{\Psi_{0,\FCI}}].
\end{equation}
On the scale of the plot, it is superimposed with the SC-FCI+LDA energy, showing that the non-self-consistent approximation is an excellent approximation for calculating the ground-state energy of the present system. The same trends have been observed in atomic and molecular systems~\cite{GinTraPraTou-JCP-21}.

We then perform linear-response calculations on-top of the ground-state SC-FCI+LDA calculations according to Eq.~(\ref{ABBASS}). The required LDA basis-set correction kernel is obtained by differentiation of Eq.~(\ref{vBLDAr})
\begin{eqnarray}
\bar{f}^\B_\LDA[\rho](\b{r},\b{r}') = \;\;\;\;\;\;\;\;\;\;\;\;\;\;\;\;\;\;\;\;\;\;\;\;\;\;\;\;\;\;\;\;\;\;\;\;\;\;\;\;\;\;\;\;\;
\nonumber\\
\left[
2 \left. \frac{\d \bar{\varepsilon}^{\mathcal{B}}(\rho)}{\d \rho} \right|_{\rho = \rho(\b{r})} + \rho(\b{r}) \left. \frac{\d^2 \bar{\varepsilon}^{\mathcal{B}}(\rho)}{\d \rho^2} \right|_{\rho = \rho(\b{r})} \right] \delta(\b{r}-\b{r}').
\label{}
\end{eqnarray}
The resulting linear-response basis-set corrected excitation energies, labelled as LR-FCI+LDA, are reported in Fig.~\ref{Fig:ExcitedStates} as a function of the basis set $n_\text{max}$ for the four excited states considered in Table~\ref{tab:exact}, and compared to the excitation energies obtained by standard FCI. The first thing to note is that the FCI excitation energies have a much faster basis convergence than the FCI ground-state energy. This is somehow expected since the same electron-electron cusp condition applies for both the ground state and the considered excited states, and therefore the short-range correlation effects normally responsible for the slow basis convergence should partially cancel out in the excitation energies. Accelerating the basis convergence of excitation energies is thus a more subtle task than accelerating the basis convergence of ground-state energies. In fact, LR-FCI-LDA does not provide any improvement over standard FCI but instead mostly deteriorates the basis convergence of excitation energies. We may attribute these disappointing results to the limited accuracy of the LDA basis-set correction potential and kernel. 

In Fig.~\ref{Fig:ExcitedStates}, we also show excitation energies obtained with a basis-set correction kernel set to zero, such that 
\begin{equation}
	A_{I,J} \simeq \bra{\Phi_I} \hat{H}^\B_\text{eff}  -{\cal E}_0^\B\ket{\Phi_J},
\end{equation}
and
\begin{equation}
B_{I,J} \simeq 0.
\end{equation}
This approximation is labelled as LR-FCI+LDA (K=0) in the figures of the present paper. It is somewhat consoling to see that the LDA kernel does nevertheless improve the excitation energies, albeit sometimes by a small amount. Finally, Fig.~\ref{Fig:ExcitedStates} also reports the excitation energies obtained by the non-self-consistent state-specific approach of Ref.~\onlinecite{GinSceTouLoo-JCP-19}, labelled as FCI+LDA. In this approach, the excited-state energy of the $n^\text{th}$ excited state is estimated as
\begin{equation}
E_{n,\text{FCI+LDA}} = E_{n,\FCI} + \bar{E}^{\B}_\LDA[\rho_{\Psi_{n,\FCI}}],
\end{equation}
where $E_{n,\FCI} = \bra{\Psi_{n,\FCI}} \hat{H} \ket{\Psi_{n,\FCI}}$ is the FCI total energy of the $n^\text{th}$ excited state with wave function $\Psi_{n,\FCI}$. The excitation energy is then given by $E_{n,\text{FCI+LDA}} - E_{0,\text{FCI+LDA}}$. Globally, the state-specific FCI+LDA approach gives excitation energies quite similar to the LR-FCI+LDA method, except for the state $(0,4)$ where FCI+LDA gives clearly excitation energies that more rapidly converge with the basis size. Since the state-specific FCI+LDA approach only involves the energy density functional $\bar{E}^{\B}_\LDA[\rho]$ and not its derivatives, it may indicate that LDA is more accurate for the energy than for the potential and kernel.

We discuss now the total excited-state energies which are reported in Fig.~\ref{Fig:ExcitedStatesTotalEnergy} as a function of the basis set $n_\text{max}$ for the four excited states considered. Here, we observe that the FCI excited-state energies exhibit a similar convergence with respect to $n_\text{max}$ as the FCI ground-state energy. This is expected since, as mentioned before, the same electron-electron cusp condition applies for both the ground state and the considered excited states. In comparison to the case of the excitation energies, there is no partial cancellation of short-range electron correlation effects, and it is thus an easier task to accelerate the basis convergence of total excited-state energies. Globally, the LR-FCI+LDA excited-state energies [Eq.~(\ref{EnB})] tend to have less basis-set incompleteness error than the standard FCI excited-state energies, and converge faster with $n_\text{max}$ to the exact energies. However, the performance of the basis correction is not uniform over all the states considered. For the state $(0,2)$, the basis-set correction is very effective in reducing the error and accelerating the basis convergence. For the states $(2,0)$ and $(2,2)$, the basis-set correction again effectively reduces the error but does not seem to significantly change the convergence rate for large $n_\text{max}$. For the state $(0,4)$, the basis-set correction only reduces the error for small $n_\text{max}$ but does not improve the standard FCI energy for $n_\text{max} \gtrsim 30$. 

Comparison with the total excited-state energies obtained with a zero basis-set correction kernel [LR-FCI+LDA (K=0)] shows again that the LDA kernel improves the basis convergence, even though the effect is small for some of the states. The state-specific FCI+LDA approach gives total excited-state energies very similar to the LR-FCI+LDA ones, except again for the state $(0,4)$ where FCI+LDA gives total excited-state energies that rapidly converge with basis size.

As a final comment, we note that the FCI total energies are of course systematically higher than the exact total energies for the ground and excited states, which makes possible a partial compensation of errors in the FCI excitation energies. By contrast, the basis-set corrected total energies converge to the exact total energies from below for the ground state and from above for the excited states, and thus the basis-set corrected excitation energies do not enjoy any compensation of errors.

\section{Conclusions}
\label{sec:conclusions}

In this work, we have extended the DFT-based basis-set correction method to the linear-response formalism, allowing one to calculate excited-state energies. We have given the general linear-response equations, as well as the more specific equations for configuration-interaction wave functions. As a proof of concept, we have applied this approach to the calculations of excited-state energies in 1D two-electron model system with harmonic potential and a Dirac-delta electron-electron interaction. The results obtained with FCI wave functions expanded in a basis of Hermite functions and a LDA basis-set correction functional within the adiabatic approximation show that the present linear-response basis-set correction method unfortunately does not help in accelerating the basis convergence of excitation energies. However, it does significantly accelerate the basis convergence of excited-state total energies.

These mixed results should now be checked on real 3D molecular systems. Possibly, for these systems, an important ingredient to add to the basis-set correction functional will be the on-top pair density. The fact that the simple non-self-consistent state-specific basis-set correction approach was found in Ref.~\onlinecite{GinSceTouLoo-JCP-19} to help accelerating the convergence of excitation energies in molecular systems gives us hope that the present linear-response basis-set correction method could be useful as well for these systems.

\section*{Acknowledgements}
This project has received funding from the European Research Council (ERC) under the European Union's Horizon 2020 research and innovation programme Grant agreement No. 810367 (EMC2).

\section*{Author Declarations}
The authors have no conflicts to disclose.

\section*{Data Availability}
The data that support the findings of this study are available from the corresponding author upon reasonable request.


\end{document}